\documentclass[11pt,letterpaper,oneside,twocolumn,final]{IEEEtran}
\usepackage[fleqn]{amsmath}
\usepackage[dvips]{graphicx}
\usepackage[noadjust]{cite}
\usepackage{balance}

\begin{document}

\title{Self-Reference Ultra-Wideband Systems}
\author{Athanasios~S.~Lioumpas  \vspace{-0.0cm}}
\maketitle

\begin{abstract}
Towards employing low complexity transceivers for signal
reception in Ultra-Wideband (UWB) systems, Transmitted Reference (TR)
and Differential TR (DTR) schemes have attracted researchers' attention. In
this letter, we introduce an alternative, less complex scheme, called Self
Reference (SR) UWB transceiver, which uses a modified replica of the
received signal itself as reference pulse, resulting in double data rates compared to TR schemes. Moreover, SR eliminates the need for
delay lines at the receiver side, which constitute a major drawback of the
conventional TR and DTR schemes, while it also requires no channel estimations, resulting in lower complexity implementations and power savings. The performance of the SR scheme is investigated in high-frequency (HF) channels, showing that it offers a better or comparable performance to that of DTR, depending on the channel conditions.
\end{abstract}


\vspace{-0.0cm}

\section{Introduction}

The optimum diversity combining scheme, in terms of
performance, is the all-Rake (ARake) receiver which resolves all multipath components (MPCs) that are often more than 100 in typical Ultra-Wideband (UWB) scenarios \cite{bib6}-\cite%
{bib8}. In order to eliminate the need for the channel estimations, Hoctor and Tomlinson
proposed a simple non-coherent Transmitted Reference (TR) scheme, which was
able to capture the energy of the multipath components \cite{bib9}. An
unmodulated known reference pulse is transmitted prior to each
data-modulated pulse within the coherence time of the channel in order for
the two pulses to experience the same channel condition. An autocorrelation
receiver is applied and the received reference signal is used as a template to demodulate the data
symbol. Hence, each data bit requires the transmission of two pulses, that
is the reference and the signal one, leading in a 50\% rate loss. Moreover,
TR systems experience a 3dB performance loss, because of the usage of a
noisy template \cite{bib11}-\cite{TR}. Another drawback is the requirement
of a delay line, which increases the implementation complexity, adds extra
power consumption and interrupts the synchronization at the reception.
Synchronization in the sub-nanosecond range is the key parameter in the
signal correlation, energy capture and data demodulation of IR signals for
UWB systems \cite{bib14}.

An alternative Differential TR (DTR) scheme was presented in \cite%
{bib15}, offering double data rates and reduced inter-symbol interference (ISI). In DTR the data are differentially modulated
using the previously sent data pulse, and hence the transmission of an extra
reference pulse is not required. However, DTR still requires the implementation of the delay line.
Moreover, due to the differential functionality, erroneous detection of a
symbol may affect the correct tracing of the next one, resulting in
performance degradation.

In this letter, we introduce a novel and simple scheme for UWB applications
with binary antipodal modulation, called Self Reference UWB (SR), which uses
as a reference signal the absolute value of the received signal multiplied
by the positive Gaussian monocycle waveform. Compared to the conventional TR
schemes which transmit two pulses for one data symbol,
SR constructs the reference pulse from the received data symbol,
resulting in double data rates (like DTR), lower complexity and saving of
computational resources without adding further power consumption. Moreover, depending on the channel conditions
the SR scheme offers comparable or better performance compared to DTR
because of the elimination of the error propagation of DTR when a data symbol is
erroneously decoded, while the absence of the delay line (implemented both in
TR and DTR schemes) reduces the complexity of the system, the power
consumption and the synchronization between the transmitter and the
receiver. The performance of the SR scheme is investigated in high-frequency (HF) channels, showing that it offers a better or comparable performance to that of DTR.
\vspace{-0.0cm}
\section{System Model and Mode of Operation}

\subsection{Channel Model}

The HF channel model (proposed by the IEEE\ 802.15.3a standardization group) is widely used in UWB research works and it is used for the evaluation of the
performance of the proposed SR scheme. The HF UWB channel model is based on the
Saleh-Valenzuela channel model \cite{saleh} and is intended to represent the
channel characteristics in the frequency range from 3.1 to 10.6 GHz \cite%
{HFchannel}. According to this model, the received signal arrive in $L$
clusters each containing $K+1$ rays.
The channel impulse response of the $i$-th realization is defined as \vspace{-0.0cm}
\begin{equation}
h_{i}(t)=X_{i}\overset{L-1}{\underset{l=0}{\sum }}\overset{K}{\underset{k=0}{%
\sum }}a_{k,l}^{i}\delta \left( t-T_{l}^{i}-\tau _{k,l}^{i}\right)
\end{equation}%
where $a_{k,l}^{i}$ is the tap weight associated with the $k$-th ray of the $%
l$-th cluster, $X_{i}$ is the log-normal shadowing and $T_{l}^{i},$ $\tau
_{k,l}^{i}$ are the cluster and ray arrival times, respectively.

\begin{figure}[t!]
\centering
\includegraphics[keepaspectratio,width=8cm, trim=0cm 1.5cm 0cm 1cm, clip=true]{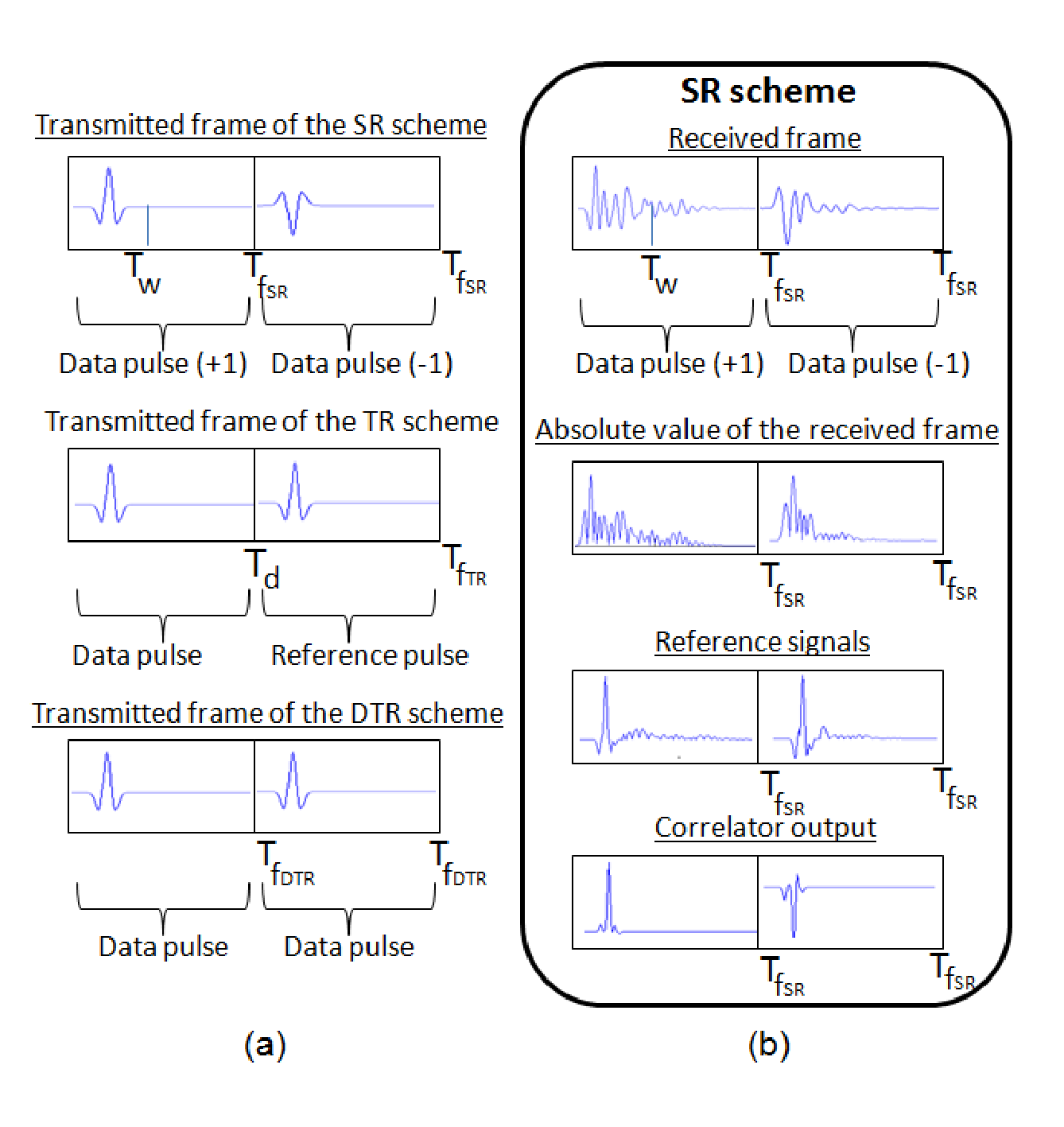}
\caption{(a) Transmitted frame of conventional TR scheme, DTR and
SR, (b) Procedure followed by data signal ($+1,-1$).} \label{Fig:Fig1}
\end{figure}
\begin{figure}[t!]
\centering
\includegraphics[keepaspectratio,width=10cm, trim=1cm 8cm 5cm 0.9cm, clip=true]{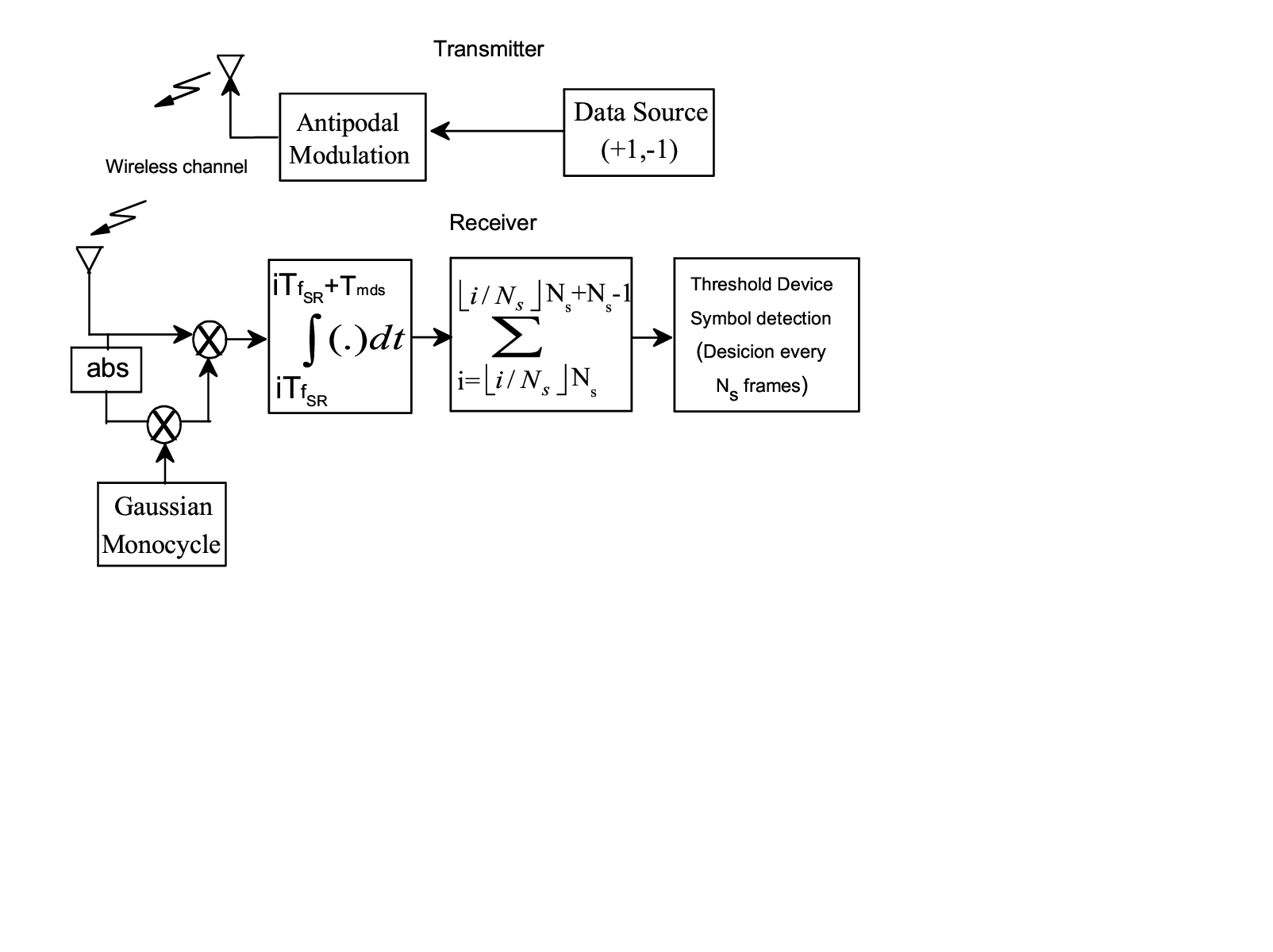}
\caption{The structure of the SR transceiver.} \label{Fig:Fig2}
\end{figure}
\vspace{-0.0cm}
\subsection{Conventional TR\ scheme}

In a binary TR UWB communication system the transmitted signal is given by
\cite{bib15}

\vspace{-0.0cm}
\begin{gather}
s_{TR}(t)= \\ \notag
\underset{s(t)}{\underbrace{\overset{\infty }{\underset{i=-\infty }%
{\sum }}b_{\lfloor i/N_{s}\rfloor }g(t-iT_{f_{TR}})}}+\underset{s_{ref}(t)}{\underbrace{\overset{\infty }{\underset{i=-\infty }{%
\sum }}g(t-iT_{f_{TR}}-T_{d})}} \label{trans}
\end{gather}%
where $b_{\lfloor i/N_{s}\rfloor }\in \left\{ 1,-1\right\} $ are the\
equiprobable data bits, $g(t)$ is the transmitted monocycle waveform that is
non-zero for $t\in (0,T_{w}),T_{w}$ is the duration of the pulse and $%
T_{f_{TR}}$ is the frame duration, which is assumed to be shorter than the
multipath delay spread $T_{mds}$ resulting in inter-pulse interference. The
index $\lfloor i/N_{s}\rfloor ,$ i.e., the integer part of $i/N_{s}$ is the
index of the data bit modulating the data waveform in the $i^{th}$ frame,
while $N_{s}$ is the number of the transmitted frames, required for
achieving an adequate bit energy at the receiver. The first sum of (2), ($s(t)$), denotes the transmitted information bits and the
second sum, ($s_{ref}(t)$), is the reference pulse transmitted $T_{d}$ seconds
later (see Fig. 1).

The transmitted signal passes through the HF multipath channel and the received
signal of the TR system can be written as \vspace{-0.0cm}
\begin{gather*}
r_{TR}(t) = \\ \notag
\underset{r(t)}{\underbrace{\overset{\infty }{\underset{%
i=-\infty }{\sum }}\overset{L-1}{\underset{l=0}{\sum }}\overset{K}{\underset{%
k=0}{\sum }}a_{k,l}^{i}b_{\lfloor i/N_{s}\rfloor }g(t-iT_{f_{TR}}-\tau
_{k,l}^{i})}}  \notag \\
 +\underset{r_{R}(t)}{\underbrace{\overset{\infty }{\underset{i=-\infty }{%
\sum }}\overset{L-1}{\underset{l=0}{\sum }}\overset{K}{\underset{k=0}{\sum }}%
a_{k,l}^{i}g(t-iT_{f_{TR}}-T_{d}-\tau _{k,l}^{i})}} + n(t)   \label{Received} \notag
\end{gather*}%
where $r(t)$ denotes the received information bits, $r_{R}(t)$ the received
reference signal and $n(t)$ represents the additive noise, which is a
zero-mean complex Gaussian random process with two-sided power spectral
density $2N_{0}$. At the receiver $r_{R}(t)$ is delayed by $T_{d}$ seconds, using delay line, and then
correlated with the first part, i.e., $r(t)$, so that a decision on the
transmitted bits can be made.

\subsection{SR scheme}

The mode of operation of the SR transceiver
is depicted in Fig. 2. Analytically, the transmitted signal of the SR is given by \vspace{-0.0cm}
\begin{equation}
s_{SR}(t)=\overset{\infty }{\underset{i=-\infty }{\sum }}b_{\lfloor
i/N_{s}\rfloor }g(t-iT_{f_{SR}})
\end{equation} where $T_{f_{SR}}$ is the frame duration of the SR, which is assumed to be shorter than $T_{mds}$
resulting in interference between data pulses. The transmitted signal passes through the HF multipath channel and is
corrupted by the AWGN. The received signal is given by
\begin{equation}
r_{SR}(t)=\overset{\infty }{\underset{i=-\infty }{\sum }}\overset{L-1}{%
\underset{l=0}{\sum }}\overset{K}{\underset{k=0}{\sum }}a_{k,l}^{i}b_{%
\lfloor i/N_{s}\rfloor }g(t-iT_{f_{SR}}-\tau _{k,l}^{i})+n\left( t\right). \notag
\vspace{0.0cm}
\end{equation}
The basic idea behind the SR scheme is that for the case of binary antipodal modulation only, we can "mimic" the positive reference pulse that is used in TR schemes, by using a modified version of the received data signal itself, e.g. the absolute value $|r_{SR}(t)|$. Considering two different cases of the HF channel model\footnote{We note that the values of Table I, are the average correlation values, over
$10^{9}$ HF\ channel realizations.}  and by simply using $|r_{SR}(t)|$\footnote{The absolute value of a signal can be obtained with a precision absolute
value circuit \cite{ABS}.} as a reference for symbol detection, the correlation between $r_{SR}(t)$ and $|r_{SR}(t)|$ is not higher than $0.4$ for the high SNR regime (see Table I). This absolute self reference scheme (ASR) is problematic due to the cutoff of the negative values of the received
signal.

\begin{table}[t!]
\caption{Correlation values assuming that the SNR$\rightarrow \infty .$}
\label{Tab1}\centering%
\begin{tabular}{|c|c|c|c|c|}
\hline
& Data signal & Ref. signal & \multicolumn{2}{|c|}{Correlation (CM1-CM2)
} \\ \hline
TR & $r(t)$ & $r_{R}(t)$ & 1 & 1 \\ \hline
ASR & $r_{SR}(t)$ & $|r_{SR}(t)|$ & 0.3829 & 0.3987 \\ \hline
SR & $r_{SR}(t)$ & $|r_{SR}(t)|g_{m}(t)$ & 0.6745 & 0.7055 \\ \hline
\end{tabular}%
\end{table}
A more precise "mimic" of the actual reference signal can be
obtained by multiplying $|r_{SR}(t)|$ with $g_{m}(t)$, where $g_{m}(t)=g(t)$ for $t\in (0,T_{w})$ and $g_{m}(t)=1$ for $t\in (T_{w}, T_{f_{SR}})$. The value of $g_{m}(t)$ for $t\ge
T_{w}$ is defined to be equal to one in order to take into consideration all
the available energy received in the duration of a frame as the transmitted
signal suffers from severe scattering because of the impact of the channel.
Consequently, this is a suboptimum solution used for collecting the spread
energy. As shown in Table I, this
multiplication almost doubles the correlation value,
which could be sufficient for making decision on the transmitted data bits.
Using the product $|r_{SR}(t)|g_{m}(t)$ as the reference signal at the receiver, a 50\% rate gain is achieved (see Fig. 1a), compared to TR
schemes.
Using this modified replica of the received signal, the correlator's output passes through a low pass filter. The energy
over $N_{s}$ frames is gathered, before a threshold device is used for
recovering the originally transmitted symbols. The SR frame is presented in Fig. 1(b) as processed at the transmitter and
the receiver, and it is shown that the frame duration, $T_{f_{SR}},$ of the
SR scheme is the half of the TR.

Compared to the conventional TR schemes, the two major advantages of the
proposed system are the absence of the transmittion of a separate reference
pulse, which doubles the data rate and the absence of a delay line at the
receiver. The implementation of a delay line at the receiver is usually the
most difficult part to implement since it must provide delay times greater
than the pulse width. The power consumption and the interruption to the
critical task of the synchronization of the non-coherent scheme are also
problems rooted by the operation of the delay line.

Compared to the DTR scheme, the SR does not require a delay line, while the
detection process of each symbol is independent of the correct detection of
the previous symbol, which may result in performance improvements, under
specific channel conditions.

\begin{figure}[]
\centering
\includegraphics[keepaspectratio,width=10cm, trim=2.0cm 1.2cm 0cm 2.5cm, clip=true]{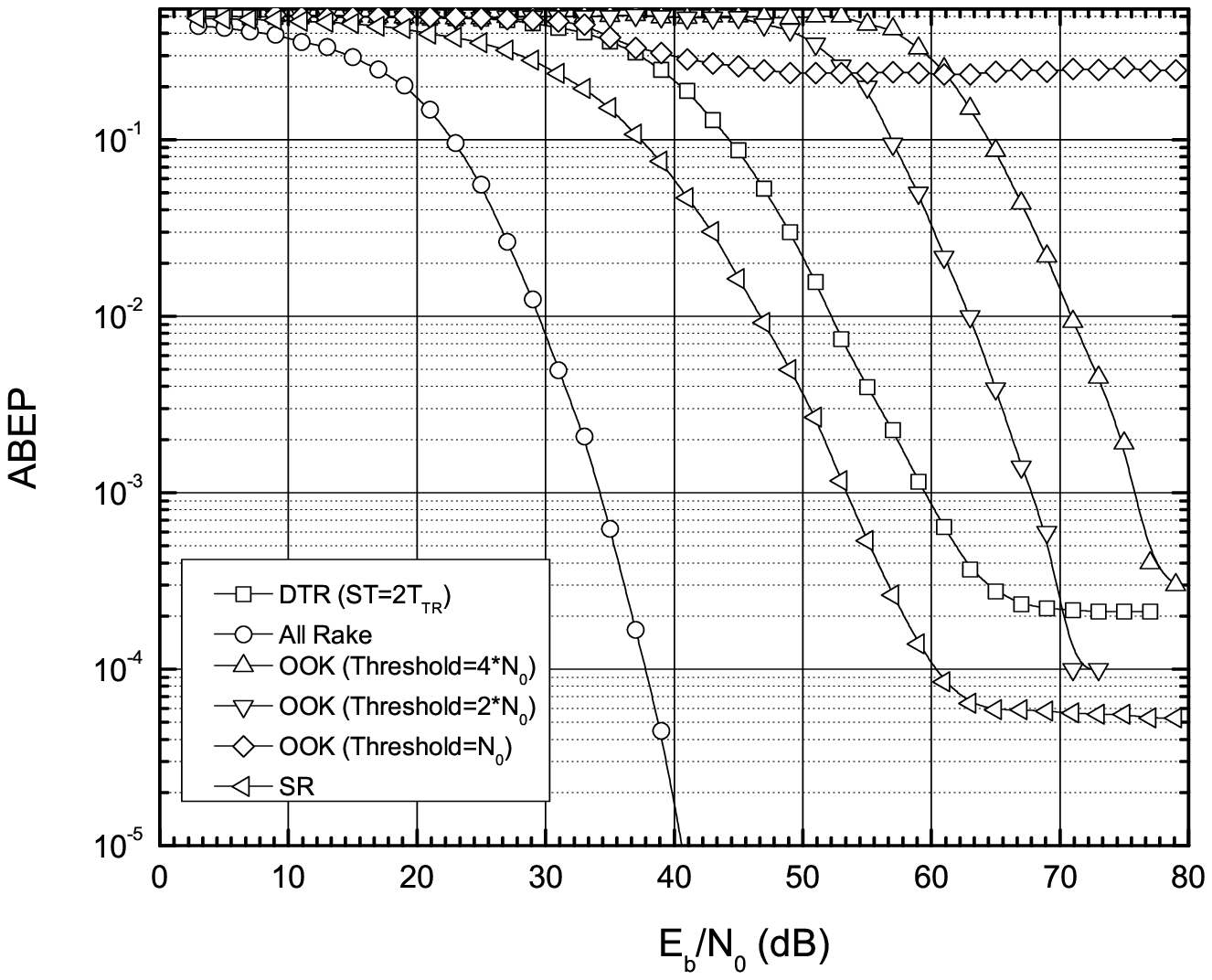}
\caption{The ABEP of the proposed SR scheme compared to DTR, the optimum ARake and the OOK schemes assuming the LOS CM1 channel model.} \label{Fig:Fig3}
\end{figure}
\vspace{-0.0cm}
\begin{figure}[]
\centering
\includegraphics[keepaspectratio,width=10cm, trim=1.2cm 1.2cm 0cm 2.5cm, clip=true]{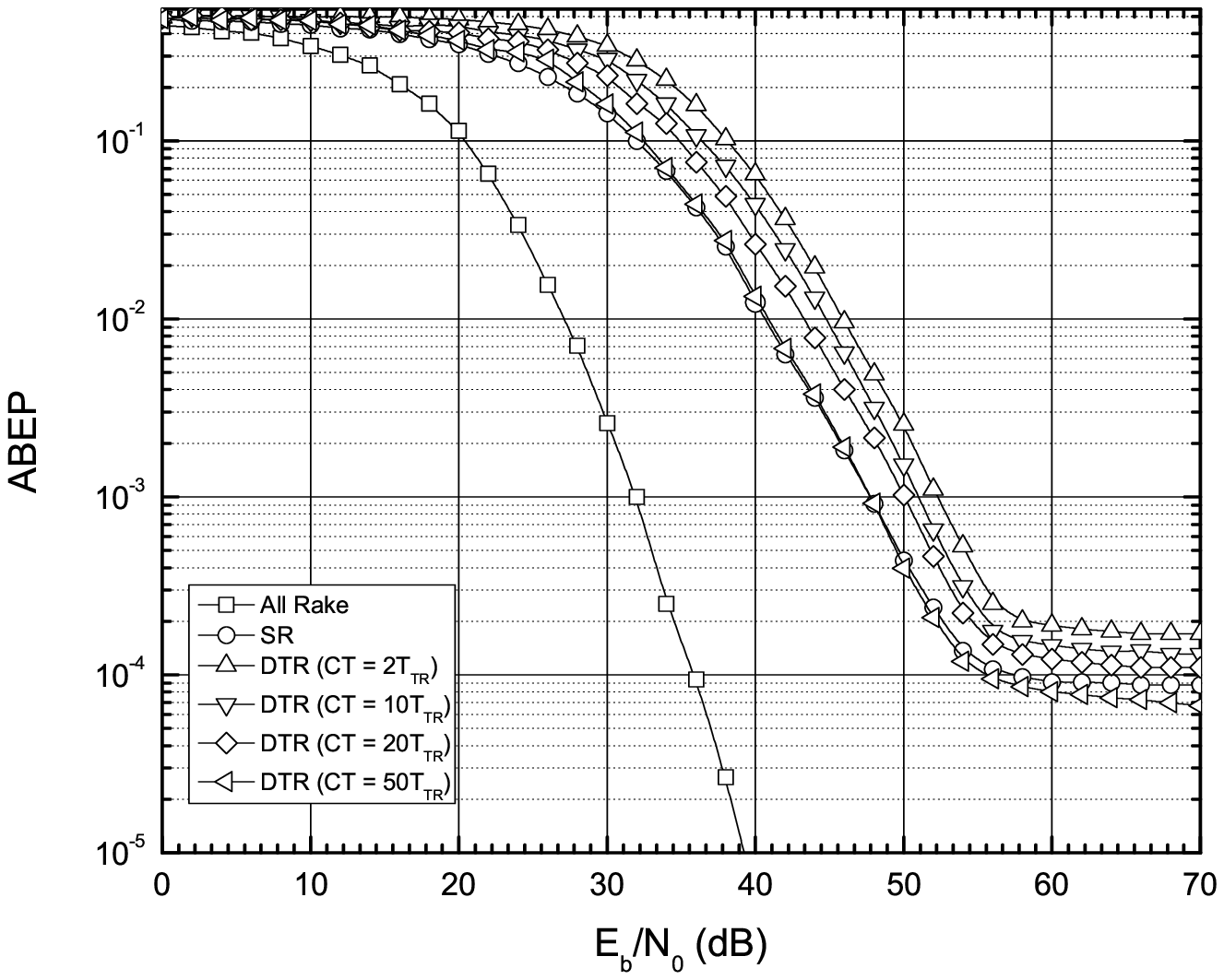}
\caption{The ABEP of the proposed SR scheme compared to DTR and the optimum ARake schemes assuming the LOS CM2 channel model and different channel coherence time.  } \label{Fig:Fig4}
\end{figure}

\section{Performance Analysis}

In this section we evaluate the performance of the proposed SR scheme in
terms of the average bit error probability (ABEP) and compare it with the
optimal All Rake, which is optimum receiver in terms of perfroamnce, the DTR and the On-Off keying (OOK) transceiver which is
the baseline receiver in terms of complexity. Performance comparisons
between the DTR and the TR\ schemes can be found in \cite{bib15}.
A second order derivative Gaussian monocycle with $T_{w}=0.7$ n$\sec $ has
been used with $T_{f_{TR}}=10.75$ n$\sec $, $T_{f_{SR}}=T_{f_{DTR}}=5.375$ n$%
\sec $ and $T_{d}=8.75$ n$\sec $ for DTR where the data pulse is delayed for
serving as reference to the next symbol. Moreover, for comparison reasons,
the simulation parameters are the same as those in \cite{bib8}, as well as
the path-loss channel model\footnote{This path loss model is the reason for the SNR range in the figures that follow. The same model is used for comparison reasons.} \cite{bib8}, \cite{bib18}. ISI is taken into
account in the analysis, while a frame per simulated symbol is assumed.

In Fig. 3, the ABEP is plotted against the Signal-to-Noise Ratio (SNR),
defined as $\frac{E_{b}}{N_{0}}$ where $E_{b}$ denotes the transmitted
symbol's energy, assuming the CM1 channel model, which refers to the line of
sight (LOS) case in residential environments \cite{bib19}. Considering the OOK scheme, and since specifying the optimum
threshold for the OOK\ receiver is beyond the scope of this paper, its
performance was examined for three different decision thresholds, i.e. N$%
_{0} $, 2N$_{0}$ and 4N$_{0}$, revealing the behavior of this receiver with
respect to that threshold. The OOK receiver's performance approaches that of
the SR's one only for high values of the SNR, where the received energy
within a time window is adequate for deciding whether a signal has been
transmitted or not.

The SR offers a comparable performance to that of DTR. This is a result of the fact that the
data detection of a symbol in DTR scheme is affected by the correct
detection of the previous one, which may result in the propagation of
errors. On the other hand, in the SR scheme the detection of each symbol is
independent from the previous one. For the case of the DTR\ and SR schemes
the error floor is caused by the ISI.

The results presented in Fig. 4 are based on CM2 channel model, which refers
to non line of sight (NLOS) case \cite{bib19}. The performance degradation of all
systems is noticeable compared to CM1 and is a consequence of the severe
scattering from which the signal suffers as we refer to a NLOS environment.
Moreover, the performance of DTR is theoretically
affected by the coherence time of the channel \cite{bib20} as the data detection in DTR is affected by the
resemblance of two consecutive channel responses. For instance, when
transmitting four symbols and assuming a channel coherence time of two
symbols, the channel response will differ between the second and the third
symbol, affecting the autocorrelation operation at the receiver and hence
the error performance. On the other hand, the SR scheme is not affected by
this assumption, since the reference pulse is directly constructed by the
same received signal. However, the performance of DTR is practically not affected, since usually the time interval between the pulses is in
the order of $100nsec$, while the coherence time is in the order of $%
0.1-1msec$.

The results are different for extreme NLOS multipath channels with
significant RMS delay spread (e.g. CM3 and CM4), where the strong multipath
components do not coincide with the first arriving ones, which results in a performance degradation for SR.
Therefore, in order to improve the performance of the proposed structure,
the duration of the transmitted pulse must be increased in order to take
advantage of the spread energy. For example, considering the CM4 channel,
a pulse duration of $T_{W}=0.1$ nsec cannot result in an ABEP lower than
0.1, while better performance can be achieved with $T_{W}=30.1$ nsec, e.g. ABEP=$10^{-3}$ for SNR=27dB.
\vspace{-0.0cm}
\section{CONCLUSIONS}

An alternative transceiver for UWB applications was introduced, called Self
Reference (SR) UWB transceiver, which uses a modified replica of the received signal itself as reference pulse, resulting in double data rates compared to TR schemes.
The proposed scheme eliminates the need for delay lines at the receiver side, as well as the need for channel estimations, resulting in lower complexity and power savings. The performance of the SR scheme was investigated in HF channels, showing that it offers a better or comparable performance to that of DTR, depending on the channel conditions.

\end{document}